\documentclass[%
preprint,
 amsmath,amssymb,
 aps,
]{revtex4-1}

\usepackage{graphicx}% Include figure files
\usepackage{dcolumn}% Align table columns on decimal point
\usepackage{bm}% bold math
\usepackage{bbold}
\begin{document}

%\preprint{APS/123-QED}

\title{Privacy amplification scheme based on composite coding}% Force line breaks with \K

\author{Wei Li$^{1,2,3}$}
\author{Shengmei Zhao$^{1,2}$}%
 \email{zhaosm@njupt.edu.cn}
\affiliation{$^{1}$Nanjing University of Posts and Telecommunications, Institute of Signal Processing and Transmission, Nanjing, 210003, China.}%
\affiliation{$^{2}$Nanjing University of Posts and Telecommunications, Key Lab Broadband Wireless Communication and Sensor Network, Ministy of Education, Nanjing, 210003, China.}%
\affiliation{$^{3}$National Laboratory of Solid State Microstructures, Nanjing University, Nanjing 210093, China.}%


\date{\today}% It is always \today, today,

\begin{abstract}

Privacy amplification is an indispensable step in the post-processing of quantum key distribution, which can be used to compress the redundancy of shared key and improve the security level of the key. The commonly used privacy amplification is based on the random selection of universal hash functions, which needs the help of an additional random source, while it does not exist in general. In this paper, we propose a privacy amplification scheme based on composite coding, which is an extension of quantum CSS codes to classical linear codes. Compared with the universal hashing function, the proposed scheme does not need other random sources, and the randomness can be completely provided by the qubit string. Furthermore, the information-theoretic bound for the extraction of the key is obvious in composite coding.

\end{abstract}

\pacs{Valid PACS appear here}% PACS, the Physics and Astronomy
\maketitle

%\tableofcontents
\section{\label{sec:level1}Introduction}

\par Quantum key distribution (QKD) is a remarkable achievement in the field of quantum information, which allows two parties, Alice and Bob, to share an unconditionally secure key for message encryption\cite{scarani2009security,diamanti2016practical,xu2020secure}. Privacy amplification is an important component in the postprocessing of practical QKDs\cite{bennett1988privacy,bennett1995generalized,deutsch1996quantum,hayashi2011exponential,hayashi2016more}. In the security research of QKD, the inevitable noise in the realistic quantum channel provides an opportunity for the eavesdropper, Eve, to implement powerful quantum side channel attacks, such as coherent attacks\cite{lo1999unconditional,shor2000simple,frohlich2017long,furrer2012continuous,sheridan2010finite}, collective attacks\cite{biham1997security,biham2002security,acin2007device,pironio2009device} and individual attacks\cite{fuchs1997optimal,bruss1998optimal,lutkenhaus2000security,waks2002security}. These channel attacks could generate correlation between Eve and Alice, and Bob, and some information of the key may be leaked to Eve. With privacy amplification, the correlation between them can be eliminated and the redundancy of the shared keys can be compressed.

\par The generalized privacy amplification is realized by means of universal hashing function, where an auxiliary random source is needed to randomly select a hashing function from the universal class\cite{bennett1995generalized,hayashi2011exponential,hayashi2016more}. The auxiliary source is called a random seed, and the privacy amplification process can be viewed as an unconditionally-secure expansion of the random seed. In general, the ideal source of randomness does not exist, one has to extract the perfect random number from an imperfect random source with the help of an $\mathit{extractor}$\cite{dodis2003extracting,nisan1996randomness,maurer1997privacy}. However, this randomness generation process will suffer a computational complexity larger than $O\left ( n\log n \right )$, with the input length of $n\geq 10^{6}$ due to the finite size effect\cite{hayashi2014security,tomamichel2012tight}. In addition, the collision probability and R$\acute{e}$nyi entropy are commonly introduced in the privacy amplification process to quantify the randomness generated by universal hashing\cite{bennett1995generalized,csiszar2011information,csiszar1978broadcast}. Due to the gap between Shannon entropy and R$\acute{e}$nyi entropy, the final key is always over compressed. Therefore, it is an interesting question that whether there exists a privacy amplification scheme that does not depend on the random selection of any universal hashing functions.

\par In the well-known security proof of QKD based on entanglement purification protocol (EPP)\cite{lo1999unconditional,shor2000simple}, error correction and privacy amplification are implemented simultaneously through Calderbank-Shor-Steane (CSS) codes\cite{bennett1996mixed,calderbank1996good,steane1996multiple}. In this kind of security proof, no universal hashing function is used, and a tight key rate can be obtained. Unfortunately, CSS codes belong to quantum error correction codes, while the postprocessings of practical QKDs are always carried out in a classical way. Are there any classical channel codes that can be used for error correction and privacy amplification simultaneously? In this paper, we propose a new scheme of privacy amplification based on classical composite linear codes. Firstly, we illustrate the relationship between error correction and information leakage in QKDs from the perspective of error correction coding. Next, we present a classical composite linear coding scheme which shares a similar structure with the CSS codes. The classical composite linear code consists of two subcodes, one of which is embedded in the other, and the two subcodes perform error correction and privacy amplification respectively.

\section{\label{sec:level1}Error correction and information leakage}

\par In the most common QKDs, such as BB84-QKD\cite{bennett1984proceedings}, E91-QKD\cite{ekert1991quantum}, measurement device independent (MDI)-QKD\cite{lo2012measurement,liu2013experimental,yin2016measurement} and twin-field (TF)-QKD\cite{lucamarini2018overcoming,ma2018phase,li2019phase}, the two communicating parties, Alice and Bob, are connected by quantum channels, and they exchange the key information through transmission of single-photon like states encoded in mutual unbiased bases (MUBs). Suppose that the eavesdropper, Eve, can do whatever she wants to do with the transmitted quantum state. In the case of ideal noise-free quantum channel, the non-cloning theorem and uncertainty principle guarantee that Eve cannot steal any key information without being detected. In fact, some inevitable quantum channel noise provides an opportunity for Eve's eavesdropping operation in theory, like intercept-resend attacks\cite{lin2011intercept}, quantum channel attacks\cite{frohlich2017long,biham1997security,fuchs1997optimal}. After all the processes before error correction, Alice, Bob and Eve each obtain a random binary bit string $R_{A}$, $R_{B}$ and $R_{E}$ with the error patterns $E_{AB}=R_{A}\oplus R_{B}$ and $E_{AE}=R_{A}\oplus R_{E}$, where $\oplus$ is the $\text{Xor}$ operation. Assume that Alice, Bob and Eve are connected by binary symmetric channels (BSC), the bits in the strings can be viewed as independent identically distributed (i.i.d) random variables.

\par In the general QKDs, Alice and Bob are in symmetrical positions. Suppose that Alice sends information to Bob in the key agreement process and the length of the transmitted qubit string is long enough, the bit error rate (BER) in the quantum channel is equal to $e_{AB}=\dfrac{d\left ( R_{A},R_{B} \right )}{n}$, the BER of Eve's bit string with respect to Alice's is $e_{AE}=\dfrac{d\left ( R_{A},R_{E} \right )}{n}$, where $d\left ( x,y \right )=W\left ( x \oplus y \right )=\sum_{i=0}^{n-1} x_{i}\oplus y_{i}$ is the Hamming distance between strings $x$ and $y$, $W\left ( z \right )$ is the weight of the codeword $z$, $n$ is the length of $R_{A}$, $R_{B}$ and $R_{E}$. According to the law of large numbers, an error pattern $E_{AB}$ is a typical sequence if its probability satisfies\cite{cover1999elements}
\begin{equation}
2^{-n\left ( H\left ( e_{AB} \right )+\epsilon \right )}\leq p\left ( E_{AB} \right )\leq 2^{-n\left ( H\left ( e_{AB} \right )-\epsilon \right )}
\end{equation}
for any $\epsilon>0$, where $H\left ( x \right )$ is Shannon entropy $H\left ( x \right )=-x\log_{2}x-\left (1-x\right )\log_{2} \left (1-x\right )$. All the typical sequences of $E_{AB}$ form a typical set $A_{\epsilon}^{AB}$ and the number of elements in $A_{\epsilon}^{AB}$ is within a range
\begin{equation}
\left ( 1-\epsilon \right )2^{n\left ( H\left ( e_{AB} \right )-\epsilon \right )}\leq \left | A_{\epsilon}^{AB} \right |\leq 2^{n\left ( H\left ( e_{AB} \right )+\epsilon \right )},
\end{equation}
in which $\epsilon$ can be infinitesimal when $n\rightarrow \infty$. Similarly, the number of typical sequences of $E_{AE}$ of $R_{E}$ with respect to $R_{A}$ is approximately $\left | A_{\epsilon}^{AE} \right | \approx 2^{n\left ( H\left ( e_{AE} \right )+\epsilon \right )}$.

\begin{figure}[ht]
\centering
\includegraphics[width=80mm]{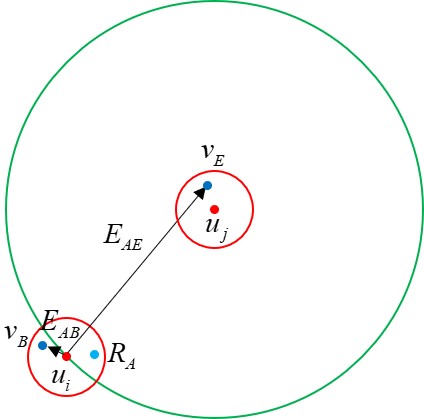}
\caption{Schematic diagram of the key agreement process.}
\label{Fig. 1}
\end{figure}

\par In the key agreement, Alice publicly share an error correction code $C$, a classical linear code $\left ( n, k\right )$, with Bob through a classical channel and it can also be passively received by Eve. Let's first review the general linear error correction coding. Alice sends a code vector $u_{ 1\times n}\in C$ over a BSC with an error probability of $p\left ( e \right )$ to Bob. For $n$ large enough, the code vector that Bob receives with a high probability is $R_{1\times n}=u_{1\times n}\oplus E_{\epsilon}$, where $E_{\epsilon}$ is a typical error pattern and $W\left ( E_{\epsilon} \right )=n p\left ( e \right )$. Let $H_{n\times \left ( n-k \right )}$ be the parity check matrix of code $C$, which consists of the bases of the dual vector space $C^{\perp}$. The corresponding syndrome cam be obtained by $S_{1\times \left ( n-k \right )}=R_{1\times n}\cdot H_{n\times \left ( n-k \right )}^{T}$, with which the typical error pattern $E_{\epsilon}$ can be computed. There are a total of $2^{n-k}$ syndromes, which means that the maximum number of error patterns that can be corrected is $2^{n-k}$. 

\par $\mathbf{Lemma\text{ } 1.}$ For any real number $t\in \left [ 0,0.5 \right ]$ and its Shannon entropy $H\left ( t \right )$,
\begin{equation}
\sum_{i=0}^{\left \lfloor nt \right \rfloor} C_{n}^{i} \leq 2^{nH\left ( t \right )},
\nonumber
\end{equation}
where $n$ is a positive number, $\left \lfloor x \right \rfloor$ is the largest integer smaller than or equal to $x$, $C_{n}^{i}$ is the combinational number formula $C_{n}^{i}=\dfrac{n!}{i!\left ( n-i\right )!}$.

\textit{Proof.} The proof of the inequality is equivalent to prove that $\sum_{i=0}^{\left \lfloor nt \right \rfloor}C_{n}^{i}2^{-nH\left ( t \right )}\leq 1$. According to the binomial theorem, we have 
\begin{equation}
\begin{split}
1=&\left [ t+\left (1-t \right) \right ]^{n}\\
=&\sum_{i=0}^{n}C_{n}^{i}t^{i}\left ( 1-t \right )^{n-i}\\
\geq & \sum_{i=0}^{\left \lfloor nt \right \rfloor}C_{n}^{i}\left ( \dfrac{t}{1-t} \right )^{i}\left ( 1-t \right )^{n}.
\end{split}
\end{equation}
As $0\leq t\leq 0.5$, therefore, $\dfrac{t}{1-t}\leq 1$ and $\left ( \dfrac{t}{1-t} \right )^{i} \geq \left ( \dfrac{t}{1-t} \right )^{nt}$ for any $i\leq \left \lfloor nt \right \rfloor$. Using these observations we see that 
\begin{equation}
\begin{split}
\sum_{i=0}^{\left \lfloor nt \right \rfloor}C_{n}^{i}\left ( \dfrac{t}{1-t} \right )^{i}\left ( 1-t \right )^{n}\geq & \sum_{i=0}^{\left \lfloor nt \right \rfloor}C_{n}^{i}\left ( \dfrac{t}{1-t} \right )^{nt}\left ( 1-t \right )^{n}\\
=&\sum_{i=0}^{\left \lfloor nt \right \rfloor}C_{n}^{i}2^{-nH\left ( t \right )},
\end{split}
\end{equation}
which completes the proof. 

\par According to Lemma 1, for any $p\left ( e \right ) \in \left [0,0.5\right ]$, we have the inequality 
\begin{equation}
\sum_{i=0}^{\left \lfloor W\left ( E_{\epsilon} \right ) \right \rfloor}C_{n}^{i}\leq 2^{n\cdot H\left ( p\left ( e \right ) \right )},
\end{equation}
where $W\left ( E_{\epsilon} \right )=np\left ( e \right )$ is the weight of the typical error pattern $E_{\epsilon}$, and according to the law of large numbers one has $\lim_{n\rightarrow \infty }\frac{W\left ( E_{\epsilon } \right )-\left \lfloor W\left ( E_{\epsilon } \right ) \right \rfloor}{n}=0$. In the Shannon limit, $nH\left ( p\left ( e \right ) \right )=n-k$. Assume that the error correction length of the code $C$ is $t=\left \lfloor W\left ( E_{\epsilon} \right ) \right \rfloor \approx W\left ( E_{\epsilon} \right )$ for sufficient large $n$, then Eq. (5) means that all the error patterns with code weight less than $t$ can all be corrected in theory. In the actual key agreement where the finite size effect should be considered, we have $n-k=fnH\left ( p\left ( e \right ) \right )$, where $f>1$ is the error correction efficiency.

\par The diagram of the key agreement is schematically illustrated shown in Fig. 1, assume that the error correction length of $C$ is $t=\left \lfloor np\left ( e_{AB} \right ) \right \rfloor +\delta$ with $\delta>0$. The red circle represents the typical set $A_{\epsilon}^{AB}$ with a Hamming radius of $t$ in the hyper space centered on the codeword $u_{i}\in C$. From Eq. (5), we can see that $\left | A_{\epsilon}^{AB} \right | \sim 2^{nH\left ( E_{AB} \right )+\epsilon}$. Alice first decodes her random string $R_{A}$ to the nearest codeword $u_{i}\in C$ and publicly announces the result $u_{i}\oplus R_{A}$ to Bob. Then Bob subtracts $u_{i}\oplus R_{A}$ from $R_{B}$, and gets $v_{B}=u_{i}\oplus E_{AB}$, where $E_{AB}=R_{A}\oplus R_{B}$. By applying the parity check matrix $H$ of code $C$ to $v_{B}$, Bob can obtain the syndrome $s_{AB}$ and calculate the error pattern $E_{AB}$, so $R_{B}$ can be corrected to $u_{i}$ as well.  Assume that the typical set $A_{\epsilon}^{AB}$ and the syndrome set $S_{AB}$ form a one-to-one mapping. For sufficiently large $n$, the probability for $A_{\epsilon}^{AB}$ satisfies $\text{Pr}\left (  A_{\epsilon}^{AB} \right )>1-\epsilon$, and the probability of occurence of a decoding error approaches 0.

\par As Bob's error correction process is conducted in private, the only thing that benefits Eve is to perform the same operations as Bob. Eve subtracts $u_{i}+R_{A}$ from $R_{E}$, and gets $v_{E}=u_{i}+E_{AE}$. With code $C$, Eve can decode $v_{E}$ to $u_{j}$, whose Haming distance from $u_{i}$ has a great probability equal to $d\left ( u_{i},u_{j} \right)\sim np\left ( e_{AE} \right)$ according to the law of large numbers. In Fig. 1, the green circle represents the typical set $A_{\epsilon}^{AE}$, within which the Hamming distance of all codevectors from $u_{j}$ is not greater than $\left \lfloor np\left ( e_{AE} \right) \right \rfloor+\delta$ with $\delta>0$. For Eve, all the codewords $u\in C$ within the green circle may equally be considered as the codeword $u_{i}$ sent from Alice to Bob. In a perfect linear code, the codewords are uniformly distributed in the codevector space. According to Eq. (5), the number of codevectors within the green circle is about $\left | A_{\epsilon}^{AE} \right |\sim 2^{nH\left ( E_{AE} \right )+\epsilon}$. Thus, the number of codewords $u\in C$ is equal to $N=\dfrac{\left | A_{\epsilon}^{AE} \right |}{\left | A_{\epsilon}^{AB} \right |}$, and Eve's probability of correctly guessing $u_{i}$ is about $\text{Pr}\approx 2^{-n\left ( H\left ( e_{AE} \right )-H\left ( e_{AB} \right ) \right )}$, and the final key rate is $r=H\left ( e_{AE} \right )-H\left ( e_{AB} \right )$. If $H\left ( e_{AE} \right )=1$, that is, Eve get no information form Alice, then $r=1-H\left ( e_{AB} \right )=\dfrac{k}{n}$. In the usual cases, the mutual information between Eve and Alice $I\left ( A;E \right)>0$, so Alice and Bob need to further compress the redundancy of the shared random string from $k$ to $n r$ with the help of privacy amplification. From Fig. 1, we can see that the optimal privacy amplification is that Alice and Bob map the codewords within the green circle one-to-one into the $\mathbb{2}^{n\cdot r}$ space.
 
\section{\label{sec:level1}Privacy amplification based on composite coding}

\par According to the Hamming distance between the codeword decoded by Eve and that decoded by Alice and Bob in the codevector space, we propose a privacy amplification scheme based on composite error correction coding. The schematic diagram of this composite coding is shown in Fig. 2. Assume that the information is sent from Alice to Bob in the key agreement, the BER of the quantum channel between Alice and Bob is $e_{AB}$ and the BER of the quantum channel between Alice and Eve is $e_{AE}$. Under the condition that Alice and Bob can extract a finite key, we have $e_{AE}>e_{AB}$ and $H\left ( e_{AE} \right )>H\left ( e_{AB} \right )$. In the composite coding, $C_{1}$ and $C_{2}$ are $\left ( n, k_{1} \right )$ and $\left ( n, k_{2} \right )$ classical linear codes with $C_{2}\subset C_{1}$, the error correction lengths of these two codes are $t_{1}=\left \lfloor n e_{AB}\right \rfloor+\delta$ and $t_{2}=\left \lfloor n e_{AE}\right \rfloor+\delta$, respectively. In Fig. 2, the codewords of $C_{2}$ are represented by red dots, the codewords of $C_{1}$ are represented by black dots, and the $i$-th codeword of $C_{2}$ is denoted as $D_{i}$. To simplify the discussion, here we assume that both $C_{1}$ and $C_{2}$ can reach the Shannon limit, then $\delta=0$, $k_{1}=n\left ( 1-H\left ( e_{AB} \right )\right )$ and $k_{2}=n\left ( 1-H\left ( e_{AB} \right ) \right )$.

\par Assume that $C_{1}$ and $C_{2}$ are perfect linear codes, their codewords are uniformly distributed in their codevector spaces. Here, we define the set $\mathbf{D_{i}}=B\left ( D_{i},t_{2} \right )$ to be
\begin{equation}
B\left ( D_{i},t_{2} \right )=\left \{ \mathbf{c} | d\left ( \mathbf{c}, D_{i}\right )\leq t_{2}, \forall \mathbf{c}\in C_{1} \right \},
\end{equation}
where $i=0,1,\cdots, 2^{k_{2}-1}$, the number elements in $\mathbf{D}_{i}$ is $N\left ( \mathbf{D}_{i} \right )=2^{n\left ( H\left ( e_{AE} \right ) -H\left ( e_{AB}\right ) \right )}$, the $a$-th elment in  $\mathbf{D}_{i}$ is denoted as $C_{i}^{a}$ with $a=0,1,\cdots, 2^{k_{1}-k_{2}}-1$, and the $0$-th element is $C_{i}^{0}=D_{i}$. Here, the allocation of indexes to the elements in $\mathbf{D}_{i}$ can be arbitrary. We first construct the set $\mathbf{D}_{0}$, where $D_{0}$ is the all zero code, and the remaining elements are the codewords whose weight is less than or equal to $t_{2}$. 

\begin{figure}[ht]
\centering
\includegraphics[width=80mm]{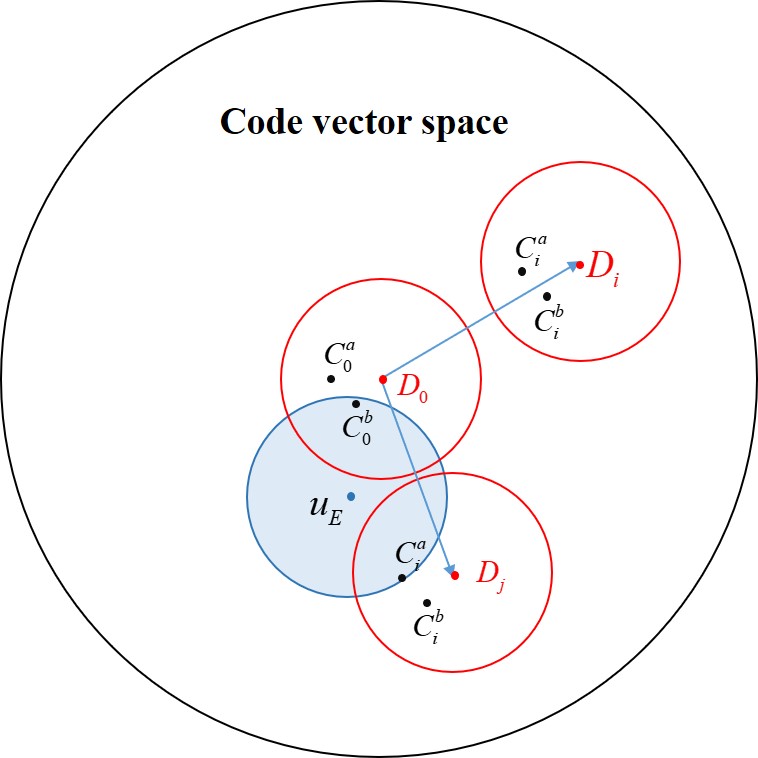}
\caption{Schematic diagram of privacy amplification based on composite coding.}
\label{Fig. 1}
\end{figure}

\par The set $\mathbf{D}_{i}$ can be obtained through $\mathbf{D}_{i}=D_{i}\oplus \mathbf{D}_{0}$, which is the coset of $\mathbf{D}_{0}$, the $b$-th codeword in $\mathbf{D}_{i}$ is $C_{i}^{b}=D_{i}\oplus C_{0}^{b}$. For any codewords $C_{i}^{a},C_{i}^{b}\in \mathbf{D}_{i}$, the Hamming distance between them is $d\left ( C_{i}^{a},C_{i}^{b} \right )\leq d\left ( C_{i}^{a},D_{i} \right )+d\left ( C_{i}^{b},D_{i} \right )\leq 2t_{2}$, which is not larger than the size of $\mathbf{D}_{i}$. However, for code $C_{2}$ whose error correction length is $t_{2}$, the Hamming distance between any two codewords satisfies $d\left ( D_{i},D_{j} \right )\geq 2t_{2}+1$. Then we have $\mathbf{D}_{i}\cap \mathbf{D}_{i}=\varnothing$, $\forall i\neq j$. Therefore, there is no intersection between $\mathbf{D}_{i}$ and $\mathbf{D}_{j}$, and this can be expressed as that for any two codewords $C_{i}^{a}\in \mathbf{D}_{i}$ and $C_{j}^{b}\in \mathbf{D}_{j}$, $C_{i}^{a}\neq C_{j}^{b}$. Another proof of this assertion is as follows. As $C_{i}^{a}\oplus C_{j}^{b}=\left ( D_{i} \oplus D_{j} \right ) \oplus \left ( C_{0}^{a} \oplus C_{0}^{b} \right )$, while $d\left ( D_{i},D_{j}\right )=W \left ( D_{i} \oplus D_{j}\right )\geq 2t_{2}+1$ and $d\left ( C_{0}^{a},C_{0}^{b}\right )=W\left ( C_{0}^{a} \oplus C_{0}^{b} \right )\leq 2t_{2}$, so $C_{i}^{a}\oplus C_{j}^{b}\neq \mathbf{0}$ is obtained, where $\mathbf{0}$ is the all zero code. Finally, we use $\text{Str}_{2}\left ( a \right )$, the equal length binary representation of the index $a$, as the privacy amplified random bit string.

\par Assume that in a practical QKD, after the key agreement between Alice and Bob, Eve decode $R_{E}$ to the codeword $u_{E}$, represented by a blue dot in Fig. 2. The blue circle is the set $\mathbf{u}_{E}=B\left ( u_{E},t_{2}\right )$, which consists of all the codewords $C\in C_{1}$ whose Hamming distance from $u_{E}$ is not larger than $t_{2}$. For Eve, the codeword obtained by Alice and Bob must not be outside the set $\mathbf{u}_{E}$. Here, $\mathbf{u}_{E}$ may intersect with several sets $\mathbf{D}_{i}$. Then we will demonstrate that the indexes of any two codewords in $\mathbf{u}_{E}$ are not equal, which is the requirement of ideal privacy amplification. Assume $\mathbf{D}_{i}$ and $\mathbf{D}_{j}$ are the two sets that intersect with $\mathbf{u}_{E}$, then the Hamming distance of any two codewords $C_{i}^{a}$ and $C_{j}^{a}$ with the same index is $d\left ( C_{i}^{a},C_{j}^{a} \right )=W\left ( C_{i}^{a}\oplus C_{j}^{a} \right )=W \left ( D_{i}\oplus D_{j} \oplus C_{0}^{a} \oplus C_{0}^{a} \right )=W \left ( D_{i}\oplus D_{j} \right )\geq 2t_{2}+1$, which is larger than the size of $\mathbf{u}_{E}$. Therefore, we can be sure that the indexes of any two elements in $\mathbf{u}_{E}$ must be different. 

\par The complete protocol for the BB84-like QKDs, which consists of BB84-QKD, E91-QKD, MDI-QKD and TF-QKD, can be expressed as follows. (0) The composite linear error correction code $\left ( n, k_{1},k_{2} \right )$ is known publicly to any parties that want to share secret key through QKD. (1) Alice and Bob choose a QKD scheme to transmit a set of quantum states randomly coded by MUBs through a quantum channel, in which each of them prepares or measures the quantum states privately. Suppose that $4n$ quantum states are successfully transmitted between Alice and Bob. (3) Alice and Bob discard the bits when they use different bases through public discussion, and each obtain a random bit string $R_{A}$ and $R_{B}$. With a high probability, the length of each bit string is $2n$. (4) Alice and Bob randomly select $n$ of these bits as check bits to evaluate the BER between them. (5) If the BER is within a predetermined value, Alice (Bob) decodes $R_{A}\left ( R_{B} \right)$ to $u_{A}$ $\left ( u_{B} \right )$ with the  $\left ( n, k_{1} \right )$ code, and sends $R_{A}\oplus u_{A}$ $\left ( R_{B}\oplus u_{B} \right )$ to Bob (Alice) through an authenticated classical channel. (6) Bob (Alice) subtracts $R_{B}$ $\left (R_{A} \right )$ from $R_{A}\oplus u_{A}$ $\left ( R_{B}\oplus u_{B} \right )$, and decode $E_{AB}\oplus u_{A} \left ( u_{B} \right )$ to  $u_{B}$ $\left ( u_{A} \right )$ with the $\left ( n, k_{1} \right )$ code, where $E_{AB}=R_{A}\oplus R_{B}$ is the error pattern between $R_{A}$ and $R_{B}$. (7) With the $\left ( n, k_{2} \right )$ code, they further decode $u_{A}$ $\left ( u_{B} \right )$ to $D_{i}$, and the probability for $u_{A}=u_{B}=u$ approaches 1 for $n\rightarrow \infty$. (8) They obtain $C_{0}^{a}$ through $C_{0}^{a}=u\oplus D_{i}$, and use $\text{Str}_{2}\left ( a \right )$, the equal length binary representation of the index of the decoded codeword, as the final shared random bit string.

\section{\label{sec:level1}Conclusion}
In this paper, We propose a privacy amplification scheme based on composite coding, and analyze the relationship between information leakage and bit error correction from the perspective of coding. Composite coding can be regarded as the extension of quantum CSS codes to classical linear codes, which integrates privacy amplification and error correction. Compared with the universal hashing function, composite coding has significant advantages in privacy amplification. For example, the randomness is only provided by the qubit string, no other random source is needed. With composite coding, the proof that the extraction of key reaches the information-theoretic bound is obvious. We anticipate that composite coding will play an important role in the post-processing of quantum information.

\section*{Acknowledgments}
This work is supported by China Postdoctoral special funding project (2020T130289), the National Natural Science Foundation of China (No. 61871234).

\bibliography{reference}

\end{document}